\documentclass[a4paper,fleqn,usenatbib]{mnras}

\usepackage{newtxtext,newtxmath}

\usepackage{graphicx}	
\usepackage{amsmath}	

\usepackage{amssymb}	

\title[TeV spectrum of Mrk 501]{On the narrow spectral feature
at $\sim$3 TeV in the MAGIC spectrum of Mrk 501}

\author[W. Hu and D. Yan]{
Wen Hu$^{1}$
and Dahai Yan$^{2}$\thanks{E-mail: yandahai@ynao.ac.cn}\\
$^{1}$Department of Physics, Jinggangshan University, Jiangxi Province, Ji'an 343009, People's Republic of China\\
$^{2}$Key Laboratory for the Structure and Evolution of Celestial Objects, Yunnan Observatories, Chinese Academy of Sciences,\\
Kunming 650011, People's Republic of China\\
}

\date{Accepted XXX. Received YYY; in original form ZZZ}

\pubyear{2021}

\begin{document}
\label{firstpage}
\pagerange{\pageref{firstpage}--\pageref{lastpage}}
\maketitle

\begin{abstract} 
Using a time-dependent one-zone leptonic model that incorporates both shock acceleration and stochastic acceleration processes, 
we investigate the formation of the narrow spectral feature at $\sim3$ TeV of Mrk 501 which was observed during the X-ray and TeV flaring activity in July 2014.
It is found that the broadband spectral energy distribution (SED) can be well interpreted as the synchrotron and synchrotron-self-Compton emission from the electron energy distribution (EED) that is composed by a power-law (PL) branch and a pileup branch. 
The PL branch produces synchrotron photons which are scattered by the electrons of the pileup branch via inverse-Compton scattering and form the narrow spectral feature observed at the TeV energies.
The EED is produced by two injection episodes, and the pileup branch in EED is caused by shock acceleration rather than stochastic acceleration.
\end{abstract}




\section{Introduction}

The broadband spectral energy distributions (SEDs) of blazars can be characterized by a double-peak structure extending from radio frequencies to the $\gamma$-ray energies \citep{Abdo2010a,Giommi2012,Acciari2021b}.
The low-energy peak located in the infrared to X-ray range is interpreted as synchrotron emission from non-thermal relativistic electrons accelerated in jets.
Based on the position of the synchrotron peak ($\nu_{\rm pk}$), blazars are divided into high-synchrotron-peaked (HSP; $\nu_{\rm pk}>10^{15}$ Hz), intermediate-synchrotron-peaked (ISP; $10^{14}<\nu_{\rm pk}<10^{15}$ Hz), and low-synchrotron-peaked (LSP; $\nu_{\rm pk}<10^{14}$ Hz) blazars \citep{Abdo2010b}.
The high-energy peak, located at $\gamma$-ray energies, can be well explained by leptonic models.
$\gamma$-ray photons from HSPs are usually  generated via inverse-Compton (IC) scattering the synchrotron photons by the same electron population that produces the synchrotron emission (SSC) \citep{Maraschi1992,Yan2014}.

Blazars exhibit fast and large-amplitude variability across the entire electromagnetic spectrum \citep[e.g.,][]{Hayashida2012,Paliya2015,Bartoli2016,Yan2018}.
However, the physical origin of the variability and the associated electron acceleration mechanism are currently not well understood. 
Diffusive shock acceleration \citep[e.g.,][]{Drury1983,Kirk1998,Summerlin2012} and stochastic acceleration \citep[e.g.,][]{Dermer1996,Katarzy2006,Yan2012} are widely considered in blazar jets.

Mrk 501 is a famous TeV HSP \citep{Quinn1996}, which is intensively monitored by various astronomical detectors, from radio frequencies to TeV $\gamma$ rays \citep[e.g.,][]{Abdo2011,Neronov2012,Shukla2015,Aleksic2015,Ahnen2018}.
Very recently, an extreme X-ray activity in July 2014 was reported by \cite{Acciari2020}. 
During the flaring activity, the X-ray and TeV emissions have been found to be correlated, 
and the amplitude of the TeV variability is twice as large as that of the X-ray variability.
 This indicates that the leptonic model is most likely favored.
Interestingly, a narrow feature at $\sim3$ TeV was observed by MAGIC stereoscopic telescope on 2014 July 19 (MJD 56857.98),
and  the significance of this peak-feature is estimated to be $\sim4\sigma$ \citep{Acciari2020}.
This kind of spectral feature is rare, which will provide insight into the underlying emission mechanism in the jet. 
\cite{Acciari2020} proposed several explanations for this narrow spectral feature, including pileup in the electron energy distribution due to
stochastic acceleration and IC pair cascade induced by electrons accelerated in a
magnetospheric vacuum gap. \cite{2021A&A...646A.115W} provided details for IC pair cascade scenario.

Interestingly, \cite{Acciari2021a} reported that the X-ray flux measured by {\emph{Swift}}-BAT in the 15-50 keV band is significantly higher 
 than that expected from the simple extrapolation of the \emph{Swift}-XRT spectral data for HSP Mrk 421 during MJD 57422-57429 (2016 February 4-11).
 This suggests an additional emission component beyond the single synchrotron emission \citep{Acciari2021a}.
This BAT excess may be related to the presence of the additional (and narrow) component in the TeV spectrum of Mrk 501 \citep{Acciari2021a}.
Actually, the results of \emph{Nu}STAR data analysis have revealed possible excess in hard X-ray emission for Mrk 421 \citep{Kataoka2016} and PKS 2155-304 \citep{Madejski2016}.
These results may imply that the appearance of the occasional TeV narrow spectral feature is not unique in Mrk 501.
The future improved TeV $\gamma$-ray detectors, such as Cherenkov Telescope Array \citep[CTA; e.g.,][]{Actis2011,Acharya2013,Sol2013}, would detect such a narrow sharp feature with high significance.
 

Our goal is to interpret the narrow feature at $\sim3$ TeV of Mrk 501 during the extreme X-ray activity in 2014 July 19 in the framework of a one-zone leptonic model.
This paper is structured as follows: we describe our model in Section \ref{sec:model}  and results in Section \ref{sec:app}, followed by the discussion and conclusions in Section \ref{sec:diss}.


\section{Model}\label{sec:model}
Our model assumes a single, homogenous emission region of radial size $R'$, 
which moves along the jet with bulk Lorentz factor $\Gamma$.
This region is assumed to filled with random magnetic field with the strength of $B'$.
The accelerated electrons in the region will radiate photons through synchrotron and inverse-Compton (IC) scattering processes.
For an observer located at a small viewing angle $\theta_{\rm obs}$ with respect to the jet, 
the radiative luminosity in the comoving frame is boosted by a factor of $\delta_{\rm D}^4$, and the variability timescale is reduced by a factor of $\delta_{\rm D}^{-1}$ \citep[e.g.,][]{Urry1995}.
Here $\delta_{\rm D}$ is the Doppler factor, which is given by $\delta_{\rm D}=2\Gamma/(1+N_\Gamma^2)$, where $N_\Gamma\equiv\Gamma\theta_{\rm obs}$ \citep{Dermer2014}.
The emitting region size is constrained by the causality condition: $R' \lesssim ct_{\rm var}\delta_{\rm D}/(1+z)$, where $t_{\rm var}$ is the measured variability timescale and $z$ is the blazar redshift.
It is usually assumed $N_\Gamma=1$, so that $\delta_{\rm D}=\Gamma$ \cite[e.g.,][]{Nalewajko2014}.

In the model, we take into account the synchrotron self-absorption, as well as the absorption for $\gamma-$ray photons caused by the interaction with the internal synchrotron radiation field.
The numerical method to calculate such emission spectra is the same as that in \cite{DM2009}.
Throughout the paper, quantities in the comoving frame of the emission region are primed, and those in the stationary frame or observer's frame are unprimed. 

In the model, we do not distinguish the acceleration region from the emission region, 
and we assume that the distribution of relativistic emitting electrons is the result of the evolution of the injected electrons undergoing acceleration, energy losses and escape.
To reduce the number of model parameters,  the injection of electrons is assumed to be mono-energetic, which is given by the $\delta-$function
\begin{eqnarray}
\dot{Q}_{\rm e,i}(\gamma',t') &= &\frac{L_{\rm inj,i}'\delta(\gamma'-\gamma_{\rm inj,i}')}{V'\gamma'_{\rm inj,i}m_{\rm e}c^2}H(t'; T_{\rm s,i}', T_{\rm f,i}') ,\ 
\end{eqnarray}
where $V'=4\pi R'^3/3$ is the comoving volume of the emission region, $m_{\rm e}$ is the rest mass of an electron,
$L_{\rm inj,i}$, and $\gamma_{\rm inj,i}'$ denote the luminosity and the Lorentz factor of the injected electrons respectively,
with the start and finish times denoted by $T_{\rm s,i}$ and $T_{\rm f,i}$, respectively.
Here, $H(x; a,b)$ denotes the Heaviside function defined by $H=1$ if $a\le x\le b$, and $H=0$ everywhere else.

\subsection{Electron Kinetic Equation} \label{seckin_eq}
After the fresh electrons are injected into the blob, 
they gain energies through accelerations and loss energies through radiative processes, and they may also escape from the blob.
The resulting electron energy distribution (EED) can be determined by solving a Fokker-Planck equation, which takes the form \citep[e.g.,][]{Park1995}
\begin{equation}
\frac{\partial N'_{\rm e}}{\partial t'} = \frac{\partial^2}{\partial \gamma'^2} \left(\frac{1}{2}\frac{d\sigma'^2}{dt'} N'_{\rm e} \right) - \frac{\partial}{\partial \gamma'}
\left(\langle\frac{d\gamma'}{dt'}\rangle N'_{\rm e} \right)  - \frac{N'_{\rm e}}{t'_{\rm esc}} + \sum_{\rm i=1}^{\rm m}\dot{Q}'_{\rm e,i},
\label{eq-transport}
\end{equation}
where the first coefficient $\frac{1}{2}\frac{d\sigma'^2}{dt'}$ acts to broaden the shape of the particle distribution, 
and the second coefficient  $\langle\frac{d\gamma'}{dt'}\rangle$ represents the mean electron acceleration rate, and $t'_{\rm esc}$ is the escape timescale. 

The term $\frac{1}{2}\frac{d\sigma'^2}{dt'}$ is represented by the diffusion coefficient, which is associated with stochastic particle-wave interactions.  
In magnetohydrodynamic (MHD) turbulence, particles can exchange energy with resonant plasma waves, 
and the associated diffusion coefficient in the energy space is given by
\begin{equation}
D_\gamma=D_0\gamma'^q,\ 
\end{equation}
where $q$ represents the spectral index of the turbulent wave spectrum, and $D_0\propto\frac{\upsilon_{\rm A}^2}{\lambda_{\rm max}^{q-1}}\left(\frac{\delta B'}{B'}\right)^2$ is dominantly determined by the Alfv\'{e}n speed ($\upsilon_{\rm A})$ and
the cutoff scale in the turbulence spectrum ($\lambda_{\rm max}$), and the ratio of the turbulence energy density in the dominant wave mode relative to the energy density of the background field ($\delta B'^2/B'^2$) \citep[e.g.,][]{Dermer1996,Stawarz2008,OSullivan2009}.

Since the properties of the turbulences are highly uncertain \citep[see][and references therein]{Teraki2019,Demidem2020}.
In this study, we use $q=2$ to simulate hard sphere scattering between the MHD waves and the electrons.
This leads the acceleration timescale independent of the energy of electrons.
 For resonant scattering by the Alfv\'{e}n waves, the index $q=2$ may be supported by the numerical simulations of freely decaying MHD turbulence \citep{Christensson2001,Brandenburg2015}. 
It should be pointed out that $q=2$ may be consistent with the nonresonant acceleration \citep{Lynn2014,Teraki2019,Demidem2020}

The mean acceleration rate is written as
\begin{equation}\label{eq: net_acc}
\langle\frac{d\gamma'}{dt'}\rangle= \dot\gamma'_{\rm sto} + \dot\gamma'_{\rm sh} + \dot\gamma'_{\rm rad}\ ,\ 
\end{equation}
where  $\dot\gamma'_{\rm sto}$, $\dot\gamma_{\rm sh}'$ and $\dot\gamma_{\rm rad}'$ denote 
the mean rate of change of the electron energy due to stochastic particle-wave interactions, shock acceleration and radiative cooling processes, respectively.
The acceleration rate due to stochastic particle-wave interactions is given by  \citep[e.g.,][]{Dermer1996,Becker2006}
 \begin{equation}
 \dot\gamma'_{\rm sto}=\frac{1}{\gamma'^2}\frac{d}{d\gamma'}\left(\gamma'^2D_\gamma\right)=4D_0\gamma',
 \end{equation}
and the acceleration rate due to repeated interactions with shock waves is given by \citep{Drury1983,Schlickeiser1984}
\begin{equation}
\dot\gamma'_{\rm sh}=A_{\rm sh}\gamma',
\end{equation}
where $A_{\rm sh}\propto \kappa u_{\rm sh}'^2/K_\parallel$ is mainly related with the volume filling factor of shock waves ($\kappa$),  the speed of shock waves moving through plasma ($u_{\rm sh}'$) and the 
spatial diffusion coefficient ($K_\parallel$ ).

To minimize the number of model parameters that are difficult to be constrained by the observed broadband SED,
 we treat $D_0$ and $A_{\rm sh}$ as physical parameters, which control the acceleration efficiency. 
Since both the shock and stochastic acceleration rates are linearly dependent on energy.
In our approach, $A_{\rm sh}$ is expressed as $A_{\rm sh}=4aD_0$, and therefore $a$ describes the relative importance of shock acceleration compared with the stochastic acceleration process.
When $a>1$, shock acceleration dominates; otherwise, stochastic acceleration dominates.
Then, the total acceleration timescale due to the incorporation of the shock and stochastic acceleration processes can be expressed as 

\begin{equation}\label{eqn:tacc1}
t'_{\rm acc}=\frac{\gamma'}{\dot\gamma'_{\rm sto}+\dot\gamma'_{\rm sh}}=\frac{1}{4D_0(1+a)}.
\end{equation}
The diffusion rate parameter $ D_0$ can be evaluated using $t'_{\rm acc}$ and $a$ by inverting the above equation to obtain $D_0=[4(1+a)t_{\rm acc}']^{-1}$,
and therefore we obtain 
\begin{equation}
\frac{1}{2}\frac{d\sigma'^2}{dt'}=\frac{\gamma'^2}{4(1+a)t_{\rm acc}'},~\dot\gamma'_{\rm sto}=\frac{\gamma'}{(1+a)t_{\rm acc}'},~\dot\gamma'_{\rm sh}=\frac{a\gamma'}{(1+a)t_{\rm acc}'}.
\end{equation}
The radiative cooling rate of the accelerated electrons due to synchrotron and SSC processes can be written as the sum $\dot\gamma'_{\rm rad}=-(b_{\rm syn}+b_{\rm ssc})\gamma'^2$, and
\begin{eqnarray}
b_{\rm syn}&=&\frac{4c\sigma_{\rm T}}{3m_ec^2}U'_{\rm B},\\
b_{\rm ssc}&=&\frac{4c\sigma_{\rm T}}{3m_ec^2}\int_0^\infty u'_{syn}(\epsilon')f_{kn}(\epsilon',\gamma')d\epsilon',
\end{eqnarray}
where $c$ is the speed of light, $\sigma_{\rm T}$ is the Thomson cross-section, $U_{\rm B}'\equiv B'^2/8\pi$ denotes the energy density of the magnetic field, 
$u_{\rm syn}'(\epsilon')$ denotes the energy density of the synchrotron radiation field,
and the function $f_{kn}(\epsilon',\gamma')$ denotes the integration of the Compton kernel \citep{Jones1968},
fully taking into account Klein-Nishina (KN) effects for an isotropic seed photon field \citep[e.g.,][]{DM2009,Hu2020}.

Using the relation between the energy-diffusion coefficient and the spatial diffusion coefficient, $D_\gamma K_\parallel=\gamma'^2\beta_{\rm A}^2c^2/9$ \citep{Schlickeiser1984,Schlickeiser1985},
the escape timescale can be evaluated as
\begin{equation}\label{eqn:tesc}
t'_{\rm esc}\simeq\frac{R'^2}{K_\parallel}=\frac{9t'^2_{\rm dyn}}{4(1+a)t'_{\rm acc}\beta_{\rm A}^2}, 
\end{equation}
with $t'_{\rm dyn} = R'/c$ and $\beta_{\rm A}$ denoting the dynamical timescale and the Alfv\'{e}n velocity normalized to the speed of light, respectively

Due to the fact that the diffusive escape velocity of electrons can not exceed the speed of light in vacuum, i.e., $t_{\rm esc}'>t'_{\rm dyn}$, 
one naturally obtains the condition
\begin{equation}\label{eqn:condit}
\frac{t'_{\rm acc}}{t'_{\rm esc}}<\frac{t'_{\rm acc}}{t'_{\rm dyn}}< \frac{9}{4(1+a)\beta_{\rm A}^2}.
\end{equation}

The Fokker-Planck equation is numerically solved through an implicit Crank-Nichelson (CN) scheme, which has the advantage of being unconditionally stable.
We use a time step $\Delta t'=0.02t'_{\rm dyn}$ \footnote{To obtain a precise result at any given time, 
the time step should be far smaller than the characteristic acceleration and the dominant energy loss timescales of the electrons radiating at the peak of the observed SED.
Using Eq. 14, we can find $t'_{\rm syn}=t'_{\rm acc}\simeq0.2t'_{\rm dyn}$ for the typical values of the parameters used in our model}.
and a 6000 point energy grid over the range $1\leq\gamma'\leq10^7$ in our code.

\subsection{Equilibrium electron spectrum for single injection}

The EED could achieve an approximate equilibrium, representing a balance between the competing processes of acceleration, cooling losses,  electron injection and escape. 
A useful quantity is the equilibrium Lorentz factor, $\gamma_{\rm eq}'$, which is determined from a balance between acceleration and cooling losses.
Because of the Klein–Nishina effect, synchrotron radiation dominates the cooling effect at $\sim\gamma_{\rm eq}'$ even if we include the SSC cooling.
Thus, we approximately have the equilibrium electron Lorentz factor 
\begin{equation}\label{eq:gameq}
\gamma_{\rm eq}'=[b_{\rm syn}t_{\rm acc}']^{-1}.
\end{equation}

Using the $\delta-$function approximation for the observed synchrotron peak frequency, $\nu_{\rm pk}=\nu_0B'\delta_{\rm D}\gamma_{\rm pk}'^2$,  
and assuming that $\gamma_{\rm eq}'$ could be approximately equal to $\gamma_{\rm pk}'$,
we find that the ratio of the acceleration timescale to the dynamical timescale is given by
\begin{eqnarray}\label{eqn:tacc}
\frac{t_{\rm acc}'}{t'_{\rm dyn}}&=&\left(t'_{\rm dyn}b_{\rm syn}\right)^{-1}\left(\frac{\nu_{\rm pk}}{\nu_0 B'\delta_{\rm D}}\right)^{-1/2}\simeq0.173(1+z)^{1/2}\\
&\times&\left(\frac{B'}{0.1~\rm G}\right)^{-\frac{3}{2}}\left(\frac{\delta_{\rm D}}{10}\right)^{-\frac{1}{2}}\left(\frac{t_{\rm var}}{1 ~\rm day}\right)^{-1}\left(\frac{\nu_{\rm pk}}{10^{18}~\rm Hz}\right)^{-\frac{1}{2}}\nonumber
\end{eqnarray}
where $\nu_0=4m_{\rm e}c^2/3hB_{\rm cr}(1+z)$ with $B_{\rm cr}\simeq4.41\times10^{13}$ G denoting the critical magnetic field strength.

The time required for establishing equilibrium is approximately evaluated through (Appendix \ref{sec:eq_time})
\begin{equation}\label{eqn:teq}
\frac{t'_{\rm eq}}{t'_{\rm dyn}} \simeq2\frac{t'_{\rm acc}}{t'_{\rm dyn}}\ln(\gamma'_{\rm eq}/\gamma'_{\rm inj}),
\end{equation}
where $\gamma'_{\rm inj}$ is the Lorentz factor of injected electrons.

In the following, we present the approximate equilibrium distributions at $t'_{\rm eq}$ for the impulsive and continual injections of mono-energetic electrons. 
For the cooling process, we take into account the synchrotron energy loss alone. 
We inject a mono-energetic electron distribution with $\gamma'_{\rm inj} =10^2$.

\subsubsection{Impulsive Injection}
\begin{figure}
\vspace{2.2mm} 
\centering
\includegraphics[width=0.45\textwidth] {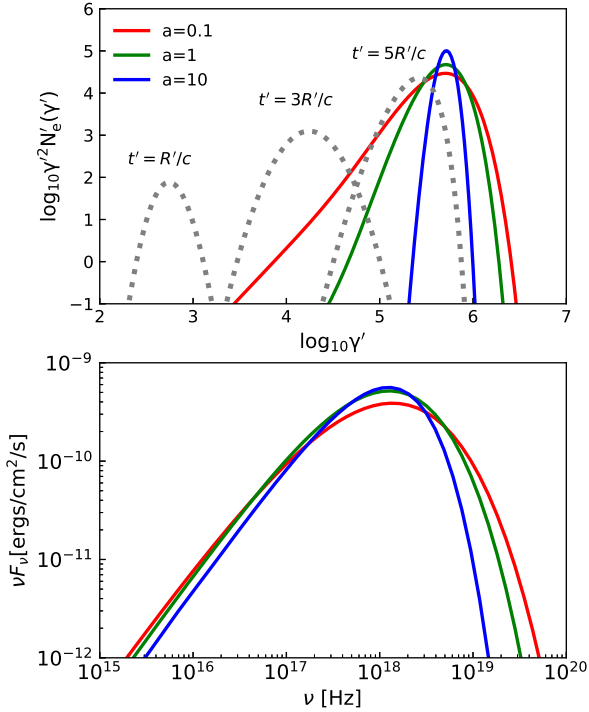}
\caption{Equilibrium EEDs (solid lines in upper panel) and the corresponding synchrotron spectra produced by the three EEDs (lower panel) in the case of impulsive injection with inefficient escape.
The evolution of EED for $a=10$ is also showed (dotted lines in upper panel). The used model parameters can be found in context.
}\label{fig:steady1}
\vspace{2.2mm}
\end{figure}
In the case of impulsive injection with inefficient escape, 
the electrons can reach the equilibrium between acceleration and energy losses. 
In the case, the electron injection luminosity, $L_{\rm inj}'$, can be related with  several observables,
which is given by (Appendix \ref{sec:Lsyn})
\begin{equation}\label{eqn:Linj1}
L_{\rm inj}'\simeq
\frac{4\pi d_L^2 F_{\rm syn}}{(\nu_{\rm pk}/\nu_0)}\frac{\gamma_{\rm inj}' B'}{\Delta T_{\rm inj}' b_{\rm syn} \delta_{\rm D}^3 }
\end{equation}
where $\Delta T_{\rm inj}'$ is the duration of injection and $F_{\rm syn}$ denotes the total flux of the synchrotron bump
\footnote{$F_{\rm syn}$ should be a factor of a few larger than the synchrotron peak flux $F_{\rm syn}^{\rm pk}$ in the $\nu-\nu F_\nu$ space. 
The accurate relation between $F_{\rm syn}$ and $F_{\rm syn}^{\rm pk}$ depends on the detail shape of the equilibrium distribution.}.

In Fig. \ref{fig:steady1}, we plot the equilibrium electron spectrum (solid lines in upper panel) 
and the corresponding synchrotron spectra produced by the three EEDs. 
Moreover, the temporal evolution of electron spectrum for $a=10$ is shown in the upper panel of Fig. \ref{fig:steady1}.
In our simulations, we set $\Delta T'_{\rm inj}=\Delta t'$ to mimic the instantaneous injection, and
the other parameters are: $B'=0.1~\rm G,~\delta_D=10, ~\beta_{\rm A}=0.1,~t_{\rm var}=0.3~\rm{day}, ~\nu_{\rm pk}=10^{18}~\rm{Hz}$ and $F_{\rm syn}=10^{-9}~\rm ergs/cm^2/s$.
For the given values of the parameters, we can obtain $\gamma'_{\rm eq}\simeq5.3\times10^5, ~L'_{\rm inj}\simeq1.5\times10^{39} ~\rm ergs/s$, $t'_{\rm acc}/t'_{\rm dyn}\simeq0.58$, $t'_{\rm eq}/t'_{\rm dyn}\simeq5$,
and $t'_{\rm esc}/t'_{\rm dyn}\simeq 350, 192, 35$ for $a=0.1,~1,~10$, respectively.

It can be seen from Fig. \ref{fig:steady1} that the width of the equilibrium distribution increases with decreasing of $a$.
When the shock acceleration is mainly responsible for the energy gain of electrons, e.g., $a=10$, 
all the electrons will be accelerated to the equilibrium energy gradually,
and a quasi-monoenergetic population of electrons is produced around the equilibrium energy.
This is different from that produced in the case of the stochastic acceleration,
where a quasi–Maxwellian distribution is formed \citep[also see][]{Katarzy2006,Tramacere2011}.
Note that the pileup distribution has been successfully employed to reproduce the hard spectra of $\gamma$-ray emission observed in several TeV blazars \citep{Lefa2011,Asano2013}.

It should be pointed out that the equilibrium distribution can be treated to be stationary as long as the equilibration timescale $t'_{\rm eq}\ll t'_{\rm esc}$.
On the other hand, we stress that the choice of $\gamma'_{\rm inj}$ cannot affect the equilibrium electron spectrum.

\subsubsection{Continuous injection}\label{sec:continu}
\begin{figure}
\vspace{2.2mm} 
\centering
\includegraphics[width=0.45\textwidth] {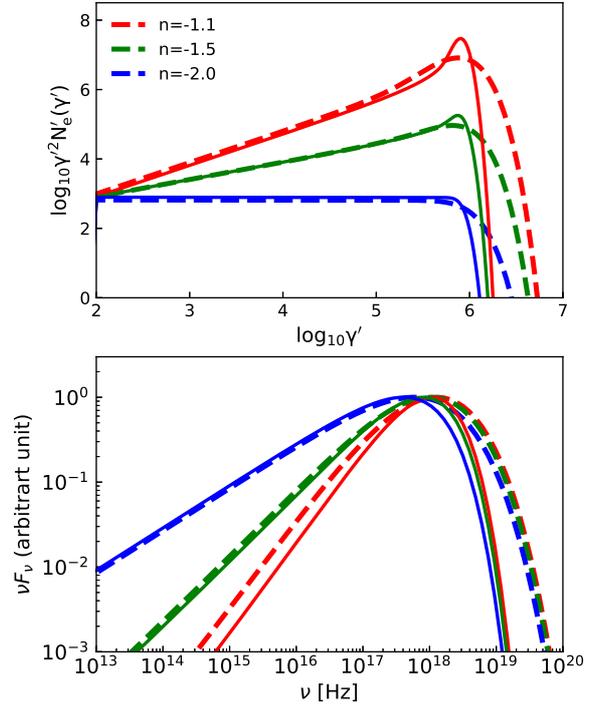}
\caption{Equilibrium EEDs (upper panel) and the corresponding synchrotron spectra produced by the EEDs (lower panel) in the case of continuous injection.
The thin solid and thick dashed lines represent $a=10$ and $a=0.1$, respectively.
}\label{fig:steady2}
\vspace{2.2mm}
\end{figure}

In the case of continuous injection,
the competition between the acceleration and the escape can produce a power-law distribution when $\gamma'\ll\gamma_{\rm eq}'$.
The power-law index of the equilibrium distribution is given by (Appendix \ref{sec:index})
\begin{equation}
n=\left(\frac{1}{2}+2a\right)-\sqrt{\left(\frac{3}{2}+2a\right)^2+4(1+a)\frac{t'_{\rm acc}}{t'_{\rm esc}}}.
\end{equation}

Therefore, we can obtain
\begin{equation}\label{eqn:slop}
\frac{t'_{\rm acc}}{t'_{\rm esc}}=\frac{n^2-(1+4a)n-(2+4a)}{4(1+a)},\ 
\end{equation}
with a given value of $a$ and $n$. Subsequently, we can calculate the Alfv\'{e}n velocity by inverting Equation \ref{eqn:tesc}, which yields
\begin{equation}\label{eqn:betaa}
\beta_A=\frac{3}{2\sqrt{1+a}}\left( \frac{t'_{\rm acc}}{t'_{\rm esc}}\right)^{1/2}\left(\frac{t'_{\rm acc}}{t'_{\rm dyn}}\right)^{-1},\ 
\end{equation}
where $t'_{\rm acc}/t'_{\rm dyn}$ is derived through Equation \ref{eqn:tacc}.
It is worth stressing that $t'_{\rm acc}/t'_{\rm dyn}$ and $t'_{\rm acc}/t'_{\rm esc}$ should satisfy the requirement of the Inequality \ref{eqn:condit}. 

In Fig. \ref{fig:steady2}, we show the equilibrium EEDs (upper panel), together with the corresponding synchrotron spectra produced by the EEDs. 
For comparison, two values of $a=0.1$ and $a=10$ are adopted in the simulations, 
and the produced synchrotron spectra are normalized to their peak flux.
To satisfy the requirement of Inequality \ref{eqn:condit} , 
the magnetic field strength $B'$ is assumed to be 0.04 G, while
 the values of $t_{\rm var}, ~\nu_{\rm pk},~\delta_{\rm D}$ are same as that in the impulsive injection. 
 We use $L'_{\rm inj}=10^{39}$ ergs/s.
With respect to the impulsive injection, we can use $n$ as a model parameter which can be estimated by the spectral index of the observed synchrotron emission.

With the given values of the parameters, 
the characteristic acceleration and equilibrium times obtained in the tests are $t'_{\rm acc}/t'_{\rm dyn}\simeq2.32$ and $t'_{\rm eq}/t'_{\rm dyn}\simeq42$, respectively.
When the stochastic acceleration dominated, i.e., $a=0.1$, we obtain $(t'_{\rm acc}/t'_{\rm esc},~\beta_{\rm A})=(1.0,0.62)$, $(0.44,0.41)$ and $(0.08,0.17)$, for $n=-2.0$, $-1.5$ and $-1.1$ respectively.
When the shock acceleration dominates, i.e., $a=10$, we obtain $(t'_{\rm acc}/t'_{\rm esc},~\beta_{\rm A})=(1.0,0.19)$, $(0.49,0.14)$ and $(0.1,0.06)$, for $n=-2.0$, $-1.5$ and $-1.1$, respectively.
These indicate that the stochastic acceleration needs more inefficient escape to provide the required slope of electron spectrum.
It is consistent with the fact the shock acceleration is more efficient than the stochastic acceleration.


One can see that in the stochastic acceleration case, the electron distribution close to the equilibrium energy can be described by a power-law turning into a log-parabola (PLLP) shape, which is consistent with previous studies \citep[e.g.,][]{Tramacere2011,Yan2012};
while, a very sharp cut-off above the equilibrium energy is formed in the shock acceleration. 
Thus, the produced synchrotron spectrum above the peak frequency drops much more rapidly than that produced by the EEDs resulting from the stochastic acceleration.


\begin{figure*}
\vspace{2.2mm} 
\centering
\includegraphics[width=\textwidth, trim={1.5cm 0 7.0cm 2.0cm},clip] {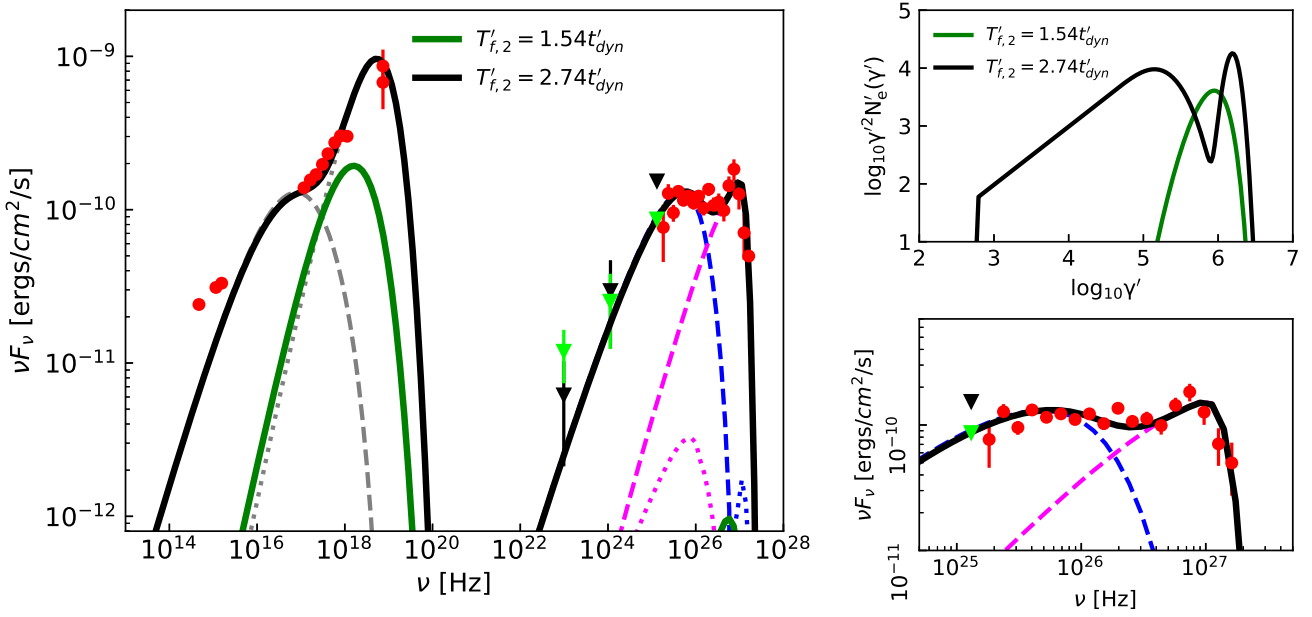}
\caption{Modeling the broadband SED of Mrk 501 during 2014 July 19 (MJD 56857.98).
The synchrotron-SSC emissions at $t'_{\rm evo}=T'_{\rm f,2}=2.74t'_{\rm dyn}$ (black solid line) and $t'_{\rm evo}=T'_{\rm f,2}=1.54t'_{\rm dyn}$ (olive line) are presented in the left panel.
The EEDs at $t'_{\rm evo}=T'_{\rm f,2}=2.74t'_{\rm dyn}$ (black solid line) and $t'_{\rm evo}=T'_{\rm f,2}=1.54t'_{\rm dyn}$ (green line) are presented in the top right panel.
In the left panel, we also present components in the SSC model at $t'_{\rm evo}=T'_{\rm f,2}=2.74t'_{\rm dyn}$. 
The gray dashed and dotted lines denote the synchrotron emission from the PL branch and pileup shape in the emitting EED, respectively.
The corresponding SSC are denoted by the blue dashed and dotted lines, respectively. 
The magenta dotted line denotes the component that the synchrotron photons produced by the electrons of the pile-up branch are scattered by the electrons of the PL branch via IC scattering.
           The magenta dashed line denotes the component that the synchrotron photons produced by the electrons of the PL branch are scattered by the electrons of the pileup branch via IC scattering.
The lower right panel shows the zoom-in view of the VHE SED. 
The \emph{Fermi}-LAT data for 4 and 10 days time intervals are denoted by the black and green triangles, respectively.
The MAGIC data from 0.1 TeV to 10 TeV have been corrected for EBL absorption according to the model of Dom\'{i}nguez et al. 2011. 
The X-ray data from \emph{Swift}-XRT and \emph{Swift}-BAT and VHE data can be safely considered simultaneous.
Details about the data can be found in Acciari et al. 2020.
}
\label{modela}
\vspace{2.2mm}
\end{figure*}

\begin{figure*}
\vspace{2.2mm} 
\centering
\includegraphics[width=\textwidth, trim={1.5cm 0 7.0cm 2.0cm},clip] {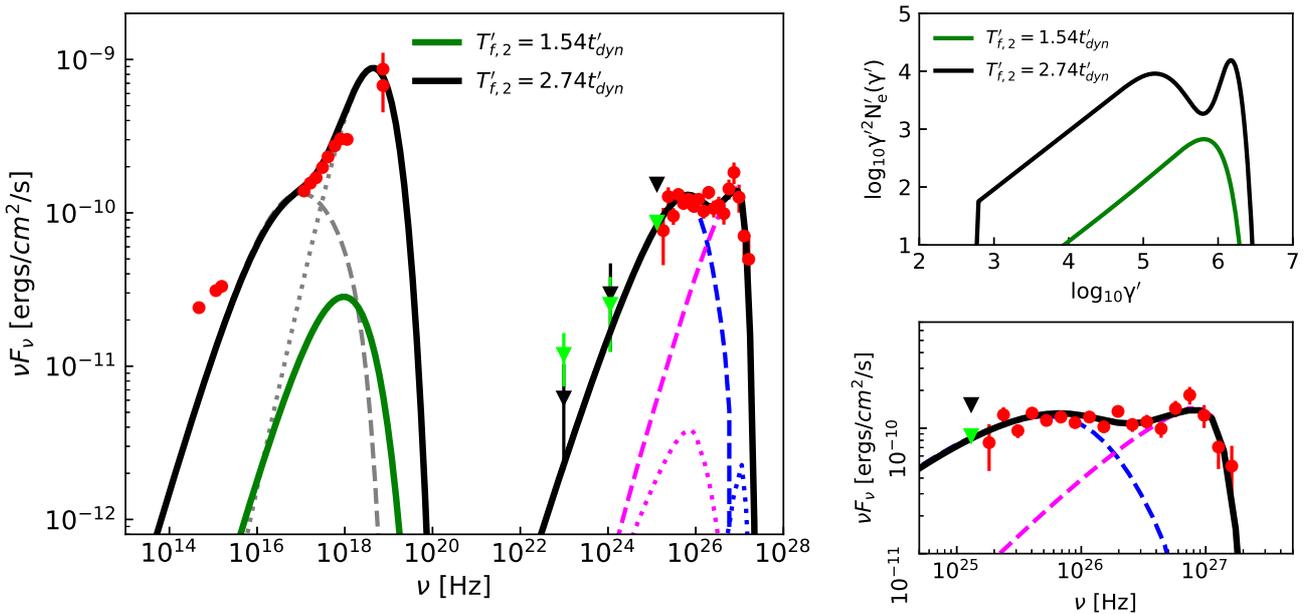}
\caption{Same as Fig.\ref{modela}, except that the first population of electrons is continuously injected.}
\label{modelb}
\vspace{2.2mm}
\end{figure*}

\section{Application to Mrk 501}\label{sec:app}
\begin{table}
	\centering
	\caption{Model parameters: the radius of the blob $R'$ is determined by the causality relation (in unit of cm); 
	the observed variability timescale $t_{\rm var}$ is in unit of day; 
	the magnetic field strength $B'$ is in unit of G; 
	the Alfv\'{e}n velocity $\beta_{\rm A}$ is in unit of the light speed;
	the diffusion coefficient $D_0$ is inferred by Eq. \ref{eqn:tacc1} (in unit of $\rm s^{-1}$); 
	the acceleration timescale $t'_{\rm acc}$ is determined by Eq. \ref{eqn:tacc};
	the peak frequency $\nu_{\rm pk,1}$ is in unit of Hz; 
	the total flux of the synchrotron bump $F_{\rm syn,1}$ is in unit of  $\rm ergs/cm^2/s$;  
	the electron injection luminosity $L'_{\rm inj, 1/2}$, in unit of ergs/s,  is derived by Eq. \ref{eqn:Linj1}; 
	$\gamma'_{\rm eq}$ and $\gamma'_{\rm pk,2}$ are the electron Lorentz factor corresponding to $\nu_{\rm pk,1}=5\times10^{18}$ Hz and $\nu_{\rm pk,2}=5\times10^{16}$ Hz, respectively.
	The equilibrium timescale $t'_{\rm eq}$ and the escape timescale $t'_{\rm esc}$ are determined by Eq. \ref{eqn:teq} and Eq. \ref{eqn:tesc}, respectively.
	Note that all the timescales are in unit of the dynamical timescale $t'_{\rm dyn}$. 
}
	\label{tab:table1}
	\begin{tabular}{lcccc} 
		\hline
		Model 			  	 & A & B & C & D  \\ 
		\hline
		$R'$ 		    			& $2.01\times10^{16}$ 	& $2.01\times10^{16}$		&$2.51\times10^{16}$	&$3.01\times10^{16}$	\\
		$\rm t_{\rm var}$ 		 & 1.0 			 	& 1.0						& 1.0					&1.0   				\\
		$\beta_{\rm A}$  		 & $0.03$   			& $0.03$ 		      			& $0.03$ 				&$0.03$ 				\\
		$a$ 			 		 & 7.0 			 	& 7.0			 	   		& 3.0					& 3.0					 \\
		$B'$ 			  	 	& $5.9\times10^{-2}$ 	& $5.9\times10^{-2}$			& $3.6\times10^{-2}$		&$2.0\times10^{-2}$    		\\ 
		$\delta_{\rm D}$ 		 & 8.0				 & 8.0						& $10.0$				&$12.0$    				\\
		$D_0$		 		&$2.4\times10^{-7}$		& $2.4\times10^{-7}$			&$2.05\times10^{-7}$	&$7.76\times10^{-8}$	\\
		$t'_{\rm acc}$ 			& $0.19$ 				& $0.19$					&$0.36$				&$0.8$			\\
		\\
		$\nu_{\rm pk,1}$ 		 & $5\times10^{18}$  	& $5\times10^{18}$			& $5\times10^{18}$		&$5\times10^{18}$		\\
		$F_{\rm syn,1}$  		  & $3\times10^{-9} $  	&$4.8\times10^{-9} $			& $3\times10^{-9} $		&$3\times10^{-9} $		\\
		$t'_{\rm eq}$ 			& $3.08$ 				& $3.08$					&$5.89$				&$13.32$			\\
		$t'_{\rm esc}$ 			& $1.6\times10^3$ 		& $1.6\times10^3$			&$1.7\times10^3$		&$7.8\times10^2$			\\
		$\gamma_{\rm eq}'$ 	   	&$1.71\times10^6$ 		& $1.71\times10^6$			&$1.96\times10^6$		&$2.40\times10^6$		\\
		$\gamma_{\rm pk,2}'$ 	& $1.71\times10^5$ 		& $1.71\times10^5$			&$1.96\times10^5$		&$2.40\times10^5$		\\
		\\
		$\gamma_{\rm inj, 1}'$ 	 & $6\times10^2$      	& $6\times10^2$ 	   		& $6\times10^2$		&$6\times10^2$		\\
		$L_{\rm inj,1}'$ 		    	& $6.58\times10^{39}$      & $1.37\times10^{38}$		&$4.43\times10^{39}$	&$3.86\times10^{39}$	\\
		$T_{\rm f,1}'$			 &$0.02$ 				& $1.54$					& 0.02				&0.02				\\
		$\gamma_{\rm inj, 2}'$ 	 & $6\times10^2$	 	& $6\times10^2$	   		& $6\times10^2$		&$6\times10^2$		\\
		$L_{\rm inj, 2}'$ 	 	 & $11.5\times10^{39}$ 	&$10.88\times10^{39}$	   	& $6.5\times10^{39}$	& $2.47\times10^{39}$	\\
		$T_{\rm s,2}'$		 	& $1.54$ 				& $1.54$					&$2.94$				&$6.66$			\\
		$T_{\rm f,2}'$			 &$2.74$ 				& $2.74$					&$5.10$				&$11.52$			\\
		\hline
	\end{tabular}
\end{table}

In this section, we explain the formation of the narrow spectral feature observed at $\sim\ $3 TeV.
We adopt a two-injection scenario.

In our model, the size of the source ($R'$), the characteristic acceleration time ($t'_{\rm acc}$) and the luminosity of the first injection electrons ($L_{\rm inj,1}'$)
could be derived from the relationships in Section \ref{sec:model} with the observed variability timescale ($t_{\rm var}$),  the synchrotron peak frequency ($\nu_{\rm pk,1}$), and the total flux of the synchrotron emission ($F_{\rm syn,1}$).
Therefore, we can describe the emission with the model parameters: $B',~\delta_{\rm D}, ~a, ~\beta_{\rm A}, ~t_{\rm var}$, $\nu_{\rm pk,1}, ~F_{\rm syn,1}$ ,$\gamma'_{\rm inj,1}, ~T'_{\rm s,1}, ~T'_{\rm f,1}$, $\gamma_{\rm inj,2}', ~L'_{\rm inj,2}$, 
$T'_{\rm s,2}$ and $T'_{\rm f,2}$.

The shortest variability found in the MWL data sample is on the order of one day \citep{Acciari2020}, we therefore use $t_{\rm var}=1$ day.
Since we consider a very inefficient electron escape from the blob, 
we fix $\beta_{\rm A}$ to 0.03, which is compatible with the quasi-linear theory \citep[e.g.,][]{Becker2006}. 
We fix $\gamma_{\rm inj,1}'=\gamma_{\rm inj,2}'=600$.
The values of $\gamma_{\rm inj,1}'$ and $\gamma_{\rm inj,2}'$ have negligible effect on the SSC emission.
We set $T'_{\rm s,1}=0$ and $T'_{\rm f,1}=\Delta t' =0.02R'/c$ for the impulsive injection.
When the first injection electrons are accelerated to be close to half of the equilibrium energy, we start the second injection.
In our test, we find that $T'_{\rm s,2}$ is not sensitive in modeling the SED.
The time interval between the two injections can be estimated by Equation \ref{time}.
The remaining parameters are free.

Fig. \ref{modela} shows SSC modeling to the observed SED of Mrk 501 during 2014 July 19 (MJD 56857.98).
The values of the parameters are given in Table \ref{tab:table1} (Model A), in which we list all input parameters and the physical quantities associated with our model. 
As shown in the figure, our modeling can reproduce the observed SED well when we select the evolution time $t'_{\rm evo}=T'_{\rm f,2}=2.74t'_{\rm dyn}$. 
We note that the optical-UV flux produced by the theoretical model is lower than the observed data.
However, it may be consistent with the variability behavior of MWL observations, which showed that the variability in the optical-UV bands is much lower than that in X-ray and TeV energies \citep{Acciari2020}. 
This indicates that the optical-UV emissions have another origin. 

We separate contributions of the different segments of the electron spectrum IC scattering different segments of the synchrotron photons.
The results indicate that the narrow TeV bump at $\sim$ 3 TeV is the contribution from the pileup bump of electrons IC scattering synchrotron photons produced by the power-law branch of electrons.
Moreover, the high flux observed by \emph{Swift}-BAT above 10 keV is interpreted as the synchrotron emission of the pileup bump of electrons which is resulting from the first injection,
while the X-rays below $\sim1$ keV is dominated by the contribution of the power-law branch of electrons resulting from the second injection.
The peak frequency of the synchrotron emission produced by the power-law branch is located at $\nu_{\rm pk,2}\simeq 5\times10^{16}$ Hz. 
With the given values of $B'$ and $\delta_{\rm D}$ reported in Table \ref{tab:table1}, one can obtain $\gamma_{\rm pk,2}'\simeq 1.7\times10^5$,
which is in good agreement with that obtained in the slow cooling model proposed by \cite{Acciari2020}.  
It should be noted that our results are obtained by using a self-consistent physical model.
Moreover, it is interesting to note that the shock acceleration rather than the stochastic acceleration is required to be dominant, i.e., a>1.

To illustrate the impact of duration of the first injection, 
we also display the result of the SED modeling with $T'_{\rm f,1}=T'_{\rm s,2}$ in Fig. \ref{modelb} with the parameters values listed in Table \ref{tab:table1} (Model B).
We have not found a significant difference between Model A and Model B.
The pileup structure in the EED can be formed in both continuous and impulsive injections.

\section{Discussion and Conclusions}\label{sec:diss}

In this work, the emitting electron distribution is determined by numerically solving a Fokker-Planck equation that self-consistently incorporates both shock and stochastic acceleration processes, radiative losses, as well as synchrotron self-absorption and internal $\gamma\gamma$ absorption.

In the model, the parameters $a$, $B'$, $\delta_{\rm D}$ and $T'_{\rm f,2}$ are four key physical quantities to reproduce the observed SED.
In the following, we examine the impact of the four parameters on SED modeling.
 
\begin{figure}
\vspace{2.2mm} 
\centering
\includegraphics[width=0.45\textwidth] {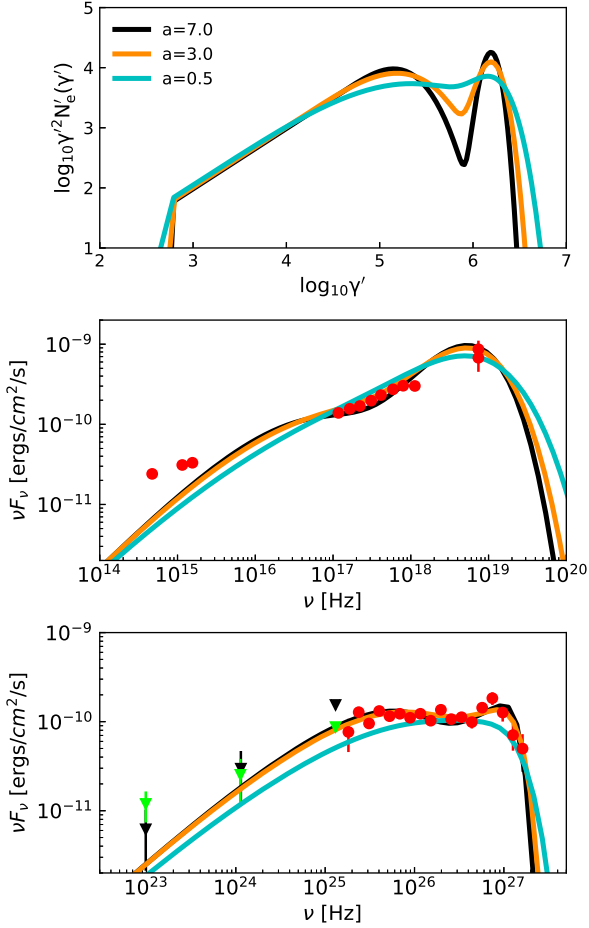}
\caption{Theoretical SEDs produced by varying the parameter $a$.
The top panel shows the EEDs. The remaining panels show the synchrotron (middle) and the SSC (bottom) spectra.
}
\label{pvar:a}
\vspace{2.2mm}
\end{figure}

In Fig. \ref{pvar:a}, we show the modeling results with varying $a$. 
The narrow feature is not visible when the stochastic acceleration dominates over the shock acceleration, i.e., $a<1$.
The necessary condition for the formation of the pile-up in electron distribution is that the electrons are well confined within the emission zone (i.e., an inefficient electron escape).
Additionally,  two injection episodes of electrons are required to reproduce the SED of Mrk 501. 
Therefore, the narrow feature in the TeV energies disappears naturally
 if the electron escape effect will be important (i.e., efficient escape scenario), and/or the first population of electrons can be neglected (i.e., a single-injection scenario). 
In those cases, the broadband SEDs reported in \cite{Acciari2020} without the sharp TeV feature should be reproduced well following the prescription in Section \ref{sec:continu}.

A constraint on the Doppler factor $\delta_D$ can be obtained from transparency of $\gamma-$rays due to pair production absorption \citep[e.g.,][]{Zdziarski1985,2008ApJ...686..181F,2012PASJ...64...80Y}.
Due to internal photon-photon annihilation, the optical depth for a $\gamma$-ray photon with observed energy $E_\gamma$ can be estimated as \citep[e.g.,][]{Abdo2011}
\begin{eqnarray}
      \tau_{\gamma\gamma} &\simeq& \frac{\sigma_TE_\gamma F_0(1+z)^2D_{\rm L}^2}{10 t_{\rm var}m_{\rm e}^2c^6\delta^6_{\rm D}} \simeq 4.65\times10^{-3}\\\nonumber
         {}& \times &\left(\frac{E_{\gamma}}{\rm TeV}\right)\left(\frac{F_0}{10^{-11} \rm erg/cm^2/s}\right)\left(\frac{t_{\rm var}}{ \rm day} \right)^{-1}\left(\frac{\delta_{\rm D}}{10}\right)^{-6} ,\ 
\end{eqnarray}
where $F_0$ is the observed monochromatic flux energy density at the observed photon energy 
\begin{equation}
E_0\simeq\frac{2\delta_{\rm D}^2m_{\rm e}^2c^4}{E_\gamma(1+z)^2}\simeq 50\left(\frac{\delta_{\rm D}}{10}\right)^2\left(\frac{E_\gamma}{\rm TeV}\right)^{-1}~\rm eV.
\end{equation}
Using $F_0\simeq3\times10^{-11}~ \rm erg/cm^2/s$ at the observed energy $E_0\simeq11.3$ eV, one has $\tau_{\gamma\gamma}(3 \rm TeV)\simeq0.16$ for the given value of $\delta_{\rm D}=8$. 
This implies that $\delta_{\rm D}$ used in Model A is close to the lower limit given by the transparency of $\gamma$ rays.

\begin{figure}
\vspace{2.2mm} 
\centering
\includegraphics[width=0.45\textwidth] {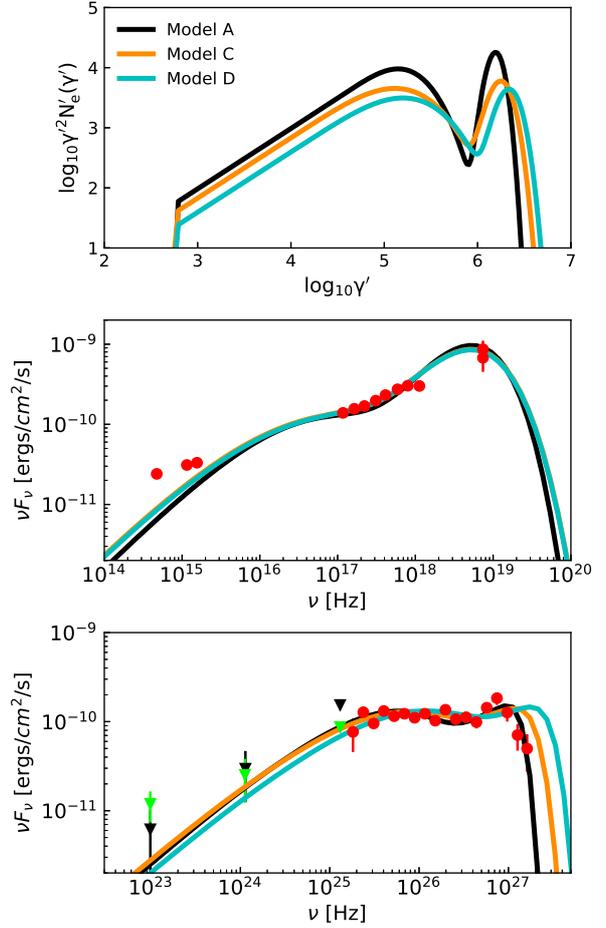}
\caption{Theoretical SEDs of Models A, C, and D.
}
\label{pvar:acd}
\vspace{2.2mm}
\end{figure}

Motivated by the above results, two typical values of $\delta_{\rm D}=10$ and 12 are used, 
and they are referred as Model C and Model D, respectively. 
In Fig. \ref{pvar:acd}, we compare the modeling results of Models A, C and D.
The parameters are given in Table \ref{tab:table1}.
We find that the location of the narrow feature moves to higher energy for higher value of $\delta_{\rm D}$, 
and a decrease of $B'$ and an increasing of $T'_{\rm f,2}$ is required to reproduce the TeV spectrum below 3 TeV.
Meanwhile, a smaller value of $a=3.0$ is also needed for matching the observed X-ray spectrum and TeV spectrum below $\sim 3$ TeV.
It is worth noting that the duration of injection $\Delta T'_{\rm inj,2} =T'_{\rm f,2}-T'_{\rm s,2}$ can be predicted from the relation
\begin{equation}
\frac{\Delta T'_{\rm inj,2}}{t'_{\rm dyn}}\simeq \frac{t'_{\rm acc}}{t'_{\rm dyn}} \ln\frac{\gamma'_{\rm pk,2}}{\gamma'_{\rm inj,2}}, 
\end{equation}
where $\gamma'_{\rm pk,2}\simeq\sqrt{\frac{\nu_{\rm pk,2}}{\nu_0B'\delta_{\rm D}}} $ is the peak frequency of the synchrotron component below 1 keV.
Using $\nu_{\rm pk,2}=5\times10^{16}$ Hz, we obtain that $\Delta T'_{\rm inj,2}/t'_{\rm dyn}$ is 1.12,  2.14 and 4.89 for Model A, C and D, respectively. 
These values are in good agreement with that obtained from the SED modeling.

We find that the global properties of the jet including $B'$, $\delta_{\rm D}$ and $R'$, as well as $\gamma_{\rm eq}'$ obtained in Model A are quite similar to that obtained by \cite{Acciari2020}.
Moreover, it can be seen from Table \ref{tab:table1} that the value of the diffusion coefficient $D_0$ is $2.4\times10^{-7}~\rm s^{-1}$. 
We find that similar result was obtained in other HBLs \citep[e.g.,][]{Lewis2016}  and Fermi bubbles \citep{Sasaki2015}.

In brief, we explain the broadband SED of Mrk 501 on 2014 July 19 by using a self-consistent one-zone leptonic jet model.
 In this model, the emitting EED is obtained by solving a Fokker-Planck equation including acceleration processes.
Two injection episodes of electrons are needed to successfully explain the SED.
 The sharp and narrow spectral feature observed at $\sim$3 TeV is resulting from the pileup branch of electrons IC scattering synchrotron photons produced by the PL branch of electrons.
 The extremely high flux observed by \emph{Swift}-BAT above 10 keV is interpreted as the synchrotron emission from the pileup bump of electrons.
  The pileup branch of electrons is the accelerated first-injection-electrons, and the PL branch of electrons is the accelerated second-injection-electrons.
Moreover, we find that shock acceleration is required to dominate over the stochastic acceleration in modeling the narrow peak feature in TeV spectrum.
This is different from the stochastic scenario proposed by \cite{Acciari2020}.

The scenario of multiple injections of relativistic electrons has been proposed to provide a explanation for 
the extreme flux variabilities of blazars \citep{Roken2009,2018A&A...616A.172R}.
\cite{2018A&A...616A.172R} argued that multiple-injection scenario is more realistic than single-injection scenario in blazar jets,
as the blazar jets could extend over parsecs to tens of kiloparsecs scales, and thus likely pick up several electron populations from intergalactic media.

In principle, the narrow feature in the VHE spectrum of Mrk 501 is expected to be detected with high significance by CTA,
which is designed to reach a sensitivity about an order of magnitude better than that of current imaging air Cherenkov telescopes and water Cherenkov telescopes \citep[e.g.,][]{Actis2011,Acharya2013,Sol2013}.
The precise measurement of the TeV narrow feature by CTA could place strong constraints on the acceleration process working in the blazar jet.

\section*{Acknowledgements}

We acknowledge the National Natural Science Foundation of China (NSFC-11803081, NSFC-12065011) 
and the joint foundation of Department of Science and Technology of Yunnan Province and Yunnan University [2018FY001(-003)]. 
The work of D. H. Yan is also supported by the CAS Youth Innovation Promotion Association and Basic research Program of Yunnan Province (202001AW070013).

\section*{data availability}
No new data were generated or analysed in support of this research.






\appendix

\section{Equilibrium Timescale}\label{sec:eq_time}
Once the values of the acceleration timescale $t'_{\rm acc}$ and the equilibrium Lorentz factor $\gamma'_{\rm eq}$ have been obtained, 
we can compute the time needed for an electron with initial Lorentz factor $\gamma_{0}'$ to be accelerated to $\gamma'$, 
by integrating the mean acceleration rate without considering the SSC energy loss, 
\begin{equation}
\frac{d\gamma'}{dt'}=-(\gamma'_{\rm eq}t'_{\rm acc})^{-1}\gamma'^2+t'^{-1}_{\rm acc}\gamma',
\end{equation}
which yields 
\begin{equation} \label{time}
\Delta T_{\rm d}'(\gamma_{0}',\gamma')= t_{\rm acc}'\ln \left(\frac{\gamma_{0}'-\gamma_{\rm eq}'}{\gamma_{0}'} \frac{\gamma'}{\gamma'-\gamma'_{\rm eq}} \right), 
\end{equation}
for $\gamma'_0\le\gamma'<\gamma_{\rm eq}'$. 
Indeed, the expression can be written in the form $\Delta T_{\rm d}'(\gamma_{0}',\gamma')\simeq t_{\rm acc}'\ln(\gamma'/\gamma'_{0})$, 
when the acceleration dominates over the radiative cooling, i.e., $\gamma_{0}'\le\gamma'\ll\gamma_{\rm eq}'$.
Moreover, we have $\Delta T_{\rm d}'(\gamma_{0}',~\gamma'=\gamma'_{\rm eq}/2)\simeq t_{\rm acc}'\ln(\gamma'_{\rm eq}/\gamma'_{\rm 0})$.

Thus, the time $t'_{\rm eq}$ required for establishing equilibrium should be at least twice as much as $\Delta T_{\rm d}'(\gamma_{0}',~\gamma'_{\rm eq}/2)$.
Here, we adopt a relatively conservative estimation $t'_{\rm eq}=2t_{\rm acc}'\ln(\gamma'_{\rm eq}/\gamma'_{0})$.

\section{Injection luminosity of electrons with inefficient escape}\label{sec:Lsyn}
The total luminosity of the synchrotron bump, neglecting self-absorption correction, can be expressed as \citep{DM2009} 
\begin{eqnarray}\label{eqn:lsyn}
L_{\rm syn} &=&m_{\rm e}c^2V'\delta_{\rm D}^4b_{\rm syn}\int_1^\infty \gamma'^2N_{\rm e}'(\gamma')d\gamma'\\\nonumber
 &=&m_ec^2V'\delta_{\rm D}^4b_{\rm syn} \langle\gamma'^2\rangle N'_{\rm tot},
\end{eqnarray}
where $\langle\gamma'^2\rangle$ is the mean value of $\gamma'^2$ and $N'_{\rm tot}$ is the total number of electrons. 

Since a quasi-Maxwellian distribution or a quasi-monoenergetic distribution 
could be expected to be formed after impulsive electron injection.
We approximately have the relation 
\begin{equation}\label{eqn:esyn}
\langle\gamma'^2\rangle\simeq\gamma_{\rm pk}'^2=\frac{\nu_{\rm pk}}{\nu_0B'\delta_{\rm D}},
\end{equation}
 where $\nu_{\rm pk}$ is the peak frequency of the observed synchrotron bump.
 
In the case of a very inefficient electron escape, one has
\begin{equation}\label{eqn:ntot}
N'_{\rm tot}\simeq\frac{L_{\rm inj}'\Delta T_{\rm inj}'}{V'\gamma_{\rm inj}' m_{\rm e}c^2} 
\end{equation}
based on the conservation of electron number.

Then, we can obtain an expression for $L'_{\rm inj}$ by combining Equations \ref{eqn:lsyn}, \ref{eqn:esyn} and \ref{eqn:ntot}, which yields
\begin{equation}
L_{\rm inj}'=\frac{L_{\rm syn}}{\Delta T_{\rm inj}' b_{\rm syn} \delta_{\rm D}^4}\frac{\gamma_{\rm inj}'}{\gamma_{\rm pk}'^2}\simeq
\frac{4\pi d_L^2 F_{\rm syn}}{(\nu_{\rm pk}/\nu_0)}\frac{\gamma_{\rm inj}' B'}{\Delta T_{\rm inj}' b_{\rm syn} \delta_{\rm D}^3 }
\end{equation}
where $\Delta T_{\rm inj}'$ denotes the duration of injection and $F_{\rm syn}$ denotes the total flux of the synchrotron bump.

\section{Approximate power-law solution}\label{sec:index}
For the continual injection of monoenergetic electrons, 
the competition between the acceleration and the escape produces a power law distribution that extends from the injected Lorentz factor up to $\gamma_{\rm eq}'$.
Following the method presented in \cite{Kroon2016}, we can obtain the power-law index of the resulting equilibrium distribution, which is in form of $N_e'(\gamma')=N_0\gamma'^n$.
By substituting the power-law form into Eq. \ref{eq-transport}, we can obtain a quadratic equation for $n$, given by
\begin{equation}
n^2-(1+4a)n-\left[2+4a+4\epsilon(1+a)\right]=0, \
\end{equation}
where $\epsilon=\frac{t'_{\rm acc}}{t'_{\rm esc}}$. 
The solution to this equation is
\begin{equation}
n_\pm=\frac{(1+4a)\pm\sqrt{(3+4a)^2+16\epsilon(1+a)}}{2}.
\end{equation}
Here, the positive power-law index $n_+$ applies below $\gamma_{\rm inj}'$, and the negative index $n_-$ applies for $\gamma'_{\rm inj}\le\gamma'\ll\gamma_{\rm eq}'$.
We find $n_-\simeq (1/2)-\sqrt{9/4+4\epsilon}$ in the case of pure stochastic acceleration, i.e., $a\rightarrow0$. 
In the case of pure shock acceleration, the power-law index $n_-$ can be written as
\begin{equation}
n_-=\left(\frac{1}{2}+2a\right)\left(1-\sqrt{1+\frac{2}{\frac{1}{2}+2a}+\frac{3\epsilon+1+2\epsilon(\frac{1}{2}+2a)}{\left(\frac{1}{2}+2a\right)^2}}\right).
\end{equation}
Making a taylor expansion in terms of the small quantity $y\equiv\left(\frac{1}{2}+2a\right)^{-1}$,
we obtain $n_-\simeq-1-\epsilon$.
This is in agreement with previous work \citep{Kirk1998}.


\bsp	
\label{lastpage}
\end{document}